\documentclass[11pt]{article}
\usepackage{geometry}
\usepackage{microtype}
\usepackage{graphicx}
\usepackage{subfigure}
\usepackage{booktabs} 
\usepackage{lscape}
\usepackage{multirow}
\usepackage[sorting=none]{biblatex}
\begin{document}

\title{Dataset of Random Relaxations for Crystal Structure Search of Li-Si System}

\author{Gowoon Cheon* \\ Stanford University\\ gcheon@stanford.edu
\and
Lusann Yang \\Google Research\\lusann@google.com

\and
Kevin McCloskey \\Google Research\\mccloskey@google.com

\and
Evan J. Reed \\Stanford University\\evanreed@stanford.edu

\and
Ekin D. Cubuk \\Google Research\\cubuk@google.com}

\maketitle

\begin{abstract}
Crystal structure search is a long-standing challenge in materials design. We present a dataset of more than 100,000 structural relaxations of potential battery anode materials from randomized structures using density functional theory calculations. We illustrate the usage of the dataset by training graph neural networks to predict structural relaxations from randomly generated structures. Our models directly predict stresses in addition to forces, which allows them to accurately simulate relaxations of both ionic positions and lattice vectors. We show that models trained on the molecular dynamics simulations fail to simulate relaxations from random structures, while training on our data leads to up to two orders of magnitude decrease in error for the same task. Our model is able to find an experimentally verified structure of a stoichiometry held out from training. We find that randomly perturbing atomic positions during training improves both the accuracy and out of domain generalization of the models. 
\end{abstract}

\section*{Introduction}
The atomic structure of a material determines its physical and chemical properties. Crystal structure search -  the prediction of unit cells that may be experimentally observed - is a long-standing problem in materials science~\cite{woodley_crystal_2008}. The number of possible configurations for a unit cell can be astronomically large even for small unit cells~\cite{oganov_structure_2019}. But among the large number of possible structures, only a few low-energy, metastable states are observed in nature. 

Most of the state-of-the-art methods for crystal structure search involve hundreds to thousands of computationally expensive ab initio calculations~\cite{noauthor_appendix_2010}, including random sampling~\cite{airss2011, airss2006}, evolutionary algorithms~\cite{lyakhov_new_2013, oganov_how_2011, tipton_grand_2013}, simulated annealing~\cite{schon_first_1996}, and basin hopping~\cite{wales_global_1999, amsler_crystal_2010}. Each of these calculations takes hours on a supercomputer, which makes them computationally challenging to use. Interatomic potentials provide faster approximations to energies and forces~\cite{zuo_performance_2020} and are a popular alternative to ab initio calculations, including recent works based on machine learning and graph neural networks~\cite{xie_crystal_2018, schutt_schnet_2018, megnet, gilmer_neural_2017, thomas_tensor_2018, schoenholz2020jax}. Gaussian approximation potentials have been proposed for a random structure search of the phosphorus system in ~\cite{deringer_data-driven_2018, deringer_hierarchically_2020}, and for a genetic algorithms search of binary systems~\cite{honrao_augmenting_2020}. However, interatomic potentials may not generalize well in describing new crystal structures they were not trained on, even for a system of a single element\cite{winczewski_interatomic_2018}. The choice of interatomic potential affects the credibility of the simulations for crystal structure search.

\begin{figure}[hbt!]
  \centering
  \includegraphics[width=6cm]{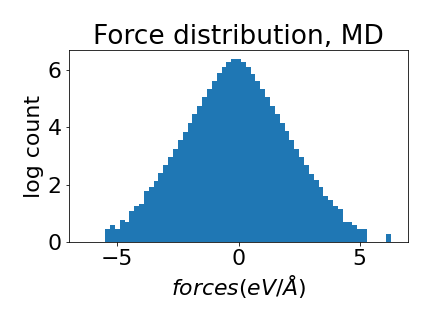}
  \includegraphics[width=6cm]{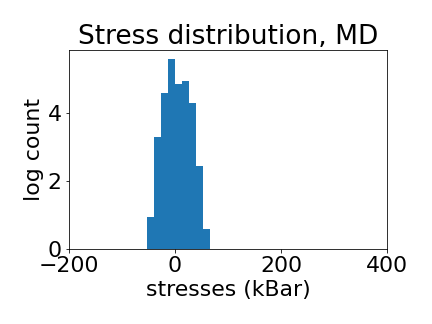}
  \includegraphics[width=6cm]{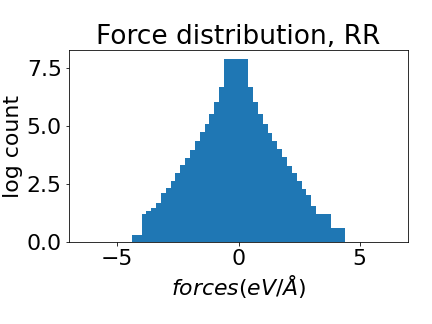}
  \includegraphics[width=6cm]{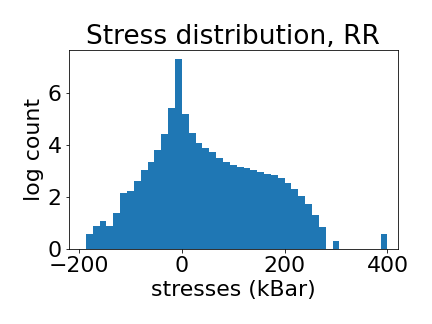}
  \caption{\textbf{Force and stress distribution of MD data(top) and random structure relaxations (RR) data(bottom).} Many interatomic potentials are fitted to MD simulation data, as MD simulations also sample from non-equilibrium structure configurations. Comparing the data from the two sampling methods shows that the distributions are different, with random relaxations having significantly higher stresses than MD simulations.}\label{data-dist}
\end{figure}

Our goal is to enable crystal structure search by training models that can cheaply relax structures from a wide variety of configurations. The creation of a machine learned potential requires training data. Most machine learned models of materials are based on molecular dynamics (MD) trajectories, which simulate the movement of atoms over time. However, structures sampled from MD trajectories are limited in the energy landscape they sample. For instance, MD simulations at lower temperatures may only sample near the equilibrium structure, while at higher temperatures they may be biased towards physical distortion mechanisms, such as sampling the phonon spectra\cite{allen_computer_1989, kong_phonon_2011}. Most datasets of materials calculations are similarly restricted to equilibrium properties ~\cite{jain_materials_2013, zagorac_recent_2019,allmann_introduction_2007,belsky_new_2002,kirklin_open_2015,choudhary_joint_2020,draxl_nomad_2019}. Sampling random structures do not suffer from the same problem, and we include all structures seen in the relaxation trajectory to ensure we sample both the non-equilibrium and equilibrium structures, similar to \cite{ocp_dataset}. In this paper, we build a dataset of structure relaxations from random initial states via high-throughput density functional theory (DFT) calculations. We show the distribution of random structures is indeed different from MD trajectories(table \ref{data-dist}) and that models trained on MD trajectories cannot predict random relaxations(section \ref{ablation}).

Our dataset includes the Li-Si system, which has at least 7 crystal structures~\cite{zeilinger_revision_2013} with different stoichiometries that have been experimentally verified~\cite{stearns_lithium_2003, dupke_structural_2012, axel_kristallstruktur_1965, schafer_kristallstruktur_1965, zeilinger_revision_2013, frank_zur_1975, evers_lisi_1997}. It is also widely studied for applications in lithium ion batteries~\cite{magasinski2010high, rszczech_nanostructured_2011, su_silicon-based_2014, cubuk_theory_2014}, machine learning based potentials~\cite{onat_implanted_2018, artrith_constructing_2018, xu_deep-learning_2020}, and for crystal structure search~\cite{morris_thermodynamically_2014, tipton_structures_2013}. Our data is easily accessible and is compatible with popular materials science packages to foster the development of data-driven methods for crystal structure search. 

To better examine the value of this dataset in a broader research context, we have prepared three related, supplementary datasets to compare the DFT relaxation trajectories to:
\begin{itemize}
    \item Molecular dynamics (MD) simulations: MD simulations are sampled to create training data for most machine learned models of materials. Compared to molecular dynamics simulations of crystalline Li-Si phases, the random initial configurations yield relaxations that cover a wider range of high-stress unit cells than observed in MD simulations.
    \item Varied DFT parameters: We include a small set of calculations using different parameters for DFT calculations, which can be used to test the out-of-domain generalizability of models trained on our data.
    \item Li-Si-O system: Li-Si-O system also has promising applications in batteries~\cite{liu_silicon_2019} but is less well-explored than the Li-Si system for crystal structure search. We include calculations from the Si-O and Li-O systems for completeness.
\end{itemize}

We demonstrate the application of this dataset to the problem of crystal structure search. We trained graph neural networks to quickly reproduce relaxation trajectories. Our networks allow us to filter randomly generated crystal structures in a fraction of the computational time used for ab initio search methods. We perform a series of experiments on models trained on the data:
\begin{itemize}
    \item Differing architectures for the prediction of the stress tensor
    \item Data augmentation using Gaussian noise
    \item Ablation experiments with different architectures, on MD data, and on energy-only trajectories
    \item Out of domain generalization to different stoichiometries and DFT parameters
    \item Application to crystal structure search
\end{itemize}

We show that our graph neural network models trained on this dataset are highly effective in simulating random relaxations for crystal structure search, with up to two orders of magnitude decrease in mean absolute error on forces and stresses compared with models trained on MD data. We improve both the accuracy and out of domain generalization, or accuracy on data that is not from the training set distribution, using data augmentation via random perturbations of atomic positions. Our models generalize well across different stoichiometries and DFT parameters.

\section*{Results}

\subsection{Dataset}
\subsubsection{Dataset of random structure relaxations}
Our random structure relaxations start with a 'random unit cell', which is constructed by choosing a set of random unit cell lengths and angles, and placing the atoms in random locations in the unit cell as implemented in AIRSS~\cite{airss2006,airss2011}. Then we compute relaxation trajectories using DFT, which are computed via a series of individual structure computations called ionic relaxation steps. Each ionic relaxation step yields accurate structures, forces, and stresses if it achieves electronic convergence. After each ionic step of DFT, the atoms and lattice are moved in the direction of the respective forces and stresses. Our dataset consists of the structures, forces, stresses, and energies at each step of the relaxation trajectory. The entire relaxation trajectory achieves structural convergence when the force difference between consecutive ionic steps are below below 10\textsuperscript{-2} eV/\AA. The Vienna Ab initio Simulation Package (VASP)\cite{kresse_software_1993} is used to relax random structures via the MedeA software environment~\cite{medea}.

Some calculations do not converge due to VASP errors which occur for a variety of reasons. We find that increasing the minimum separation between atoms decreases the number of jobs that fail. Models trained only on trajectories that converge have the potential to incur a sampling bias that could lead to incorrect predictions for random structures that would not converge with VASP. Hence, we include trajectories from jobs that failed due to VASP errors in the dataset, indicating  which jobs did not reach electronic convergence. We also find that some relaxation trajectories with larger unit cells need more ionic relaxation steps than the maximum number of ionic steps we allowed VASP to achieve structural convergence. For these jobs, each individual step achieved electronic convergence. The intermediate structures in these trajectories may still be used for learning the energy landscape. We include these trajectories in the data, indicating which jobs did not reach ionic convergence. The number of such jobs are in our data summary in table \ref{data_summary}.

A summary of our random structure relaxation dataset can be found in table~\ref{data_summary}. There are a total of 116,200  calculations which required 55.6 million core-hours to generate. Our dataset is accompanied by Colab notebooks in the supporting information. We provide structural details such as space groups, lattice parameters and visualizations, as well as code for accessing and using the dataset. 

\subsubsection{Dataset of molecular dynamics simulations} 
We use the experimentally observed crystal structures of LiSi, Li\textsubscript{2}Si, Li\textsubscript{7}Si\textsubscript{2}, Li\textsubscript{13}Si\textsubscript{4} and Li\textsubscript{15}Si\textsubscript{4} from the Materials Project and perform molecular dynamics(MD) simulations with VASP.  We created 64 to 78-atom supercells of the conventional unit cell to ensure the unit cells were sufficiently large for molecular dynamics, as described in table \ref{lisi_data}.

\subsubsection{Different DFT parameters}
We create a dataset for studying out-of-domain generalization by sampling 800 Li\textsubscript{15}Si\textsubscript{4} structures in the dataset at random and changing the calculation parameters for DFT. The best settings for DFT will depend on the application under consideration. For high-throughput databases such as ours, parameters that determine the computational cost such as k-space sampling density may be set to coarse settings to facilitate data generation. This means that models trained on different DFT datasets are likely to be trained on different DFT parameters and may not generalize well. We enable the study of this problem by including systematically generated out-of-domain calculations in our dataset.

We change:
\begin{itemize}
    \item the k-point mesh spacing (0.5\AA\textsuperscript{-1}, 1\AA\textsuperscript{-1}; default 0.25\AA\textsuperscript{-1}). This controls how the wavefunctions are sampled in the frequency domain.
    \item the pseudopotential for lithium and the electronic minimization algorithm (Li, conjugate gradient; default Li\_sv, Normal). The choice of pseudopotential determines how the effective potential is approximated for each element, and electronic minimization algorithm is the algorithm used to minimize the electronic density.
    \item And the step size for ionic relaxation (VASP parameter POTIM, 0.8, default 0.4)
\end{itemize}

\begin{figure*}[t]
  \centering
  \includegraphics[width=0.4\linewidth]{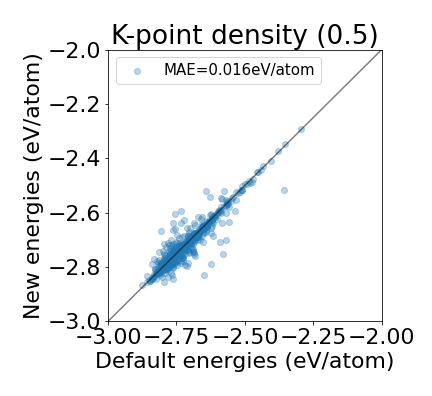} 
  \includegraphics[width=0.4\linewidth]{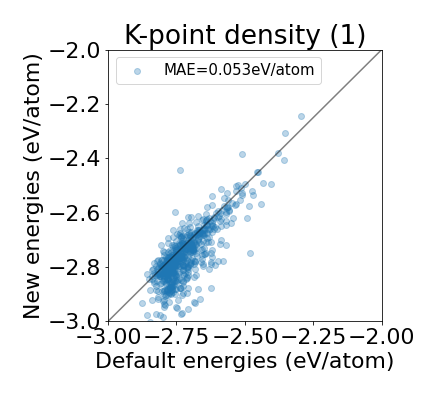} 
  \includegraphics[width=0.4\linewidth]{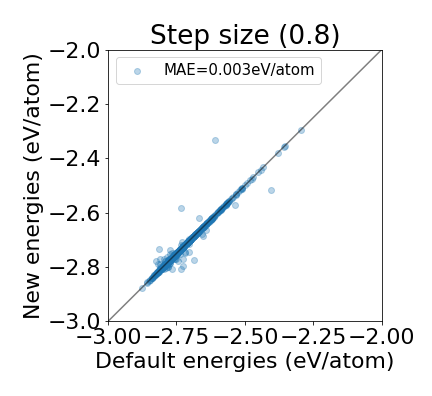} 
  \includegraphics[width=0.4\linewidth]{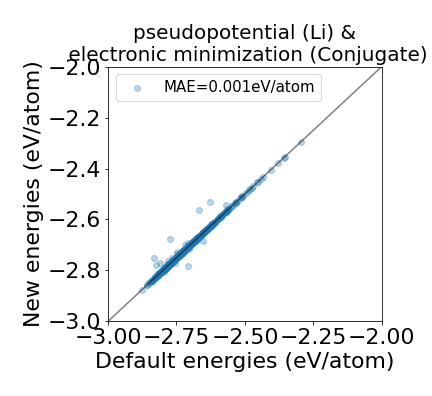} 
  \caption{\textbf{Energies of final relaxed structures under different DFT calculation parameters (y-axis), plotted with the calculations done with default values (x-axis).} We create a dataset for studying out-of-domain generalization by sampling 800 Li\textsubscript{15}Si\textsubscript{4} structures in the dataset at random and changing the calculation parameters for DFT. \\
  Top left: the k-point mesh spacing is changed from 0.25\AA\textsuperscript{-1} to 0.5\AA\textsuperscript{-1}. \\
  Top Right: the k-point mesh spacing is changed from 0.25\AA\textsuperscript{-1} to 1\AA\textsuperscript{-1}. \\
  Bottom left: the pseudopotential for lithium and the electronic minimization algorithm is changed from Li\_sv and Normal to Li and conjugate gradient.\\
  Bottom right: the step size for ionic relaxation (VASP parameter POTIM) is changed from 0.4 to 0.8.}\label{OOD_data}
\end{figure*}

We compare the energies of final relaxed structures under the changed calculation parameters with the calculations done using the dataset's default values in figure~\ref{OOD_data}.

Changing the ionic relaxation step size, the Li pseudopotential, and the electronic minimization algorithms incur small deviations in the final energies of relaxed random structures. Changing the k-point density, however, significantly affects the final energies, with an MAE of 0.016 eV/atom difference for k-point spacing 0.5\AA\textsuperscript{-1} and 0.053 eV/atom for 1\AA\textsuperscript{-1}. The data generated with k-point spacing 1\AA\textsuperscript{-1} tends to underestimate the final energies of the relaxed structures.

\subsubsection{Li-Si-O system}
We include two different stoichiometries in the Li-Si-O system that have experimentally reported structures, Li\textsubscript{4}SiO\textsubscript{4} and Li\textsubscript{2}Si\textsubscript{2}O\textsubscript{5}. We generated 5000 random relaxations of each stoichiometry. For completeness, we also include data on Li-O and Si-O stoichiometries with experimental structures as well, LiO, Li\textsubscript{2}O and SiO\textsubscript{2}. We generated 5000 random relaxations of LiO and Li\textsubscript{2}O. The SiO\textsubscript{2} is known to have many structurally different phases with different sizes, so we generated numerous random structures with 1 to 16 formula units per unit cell and performed 33000 random relaxations.

\subsubsection{Distribution forces and stresses in random structure relaxations vs MD simulations}
Many interatomic potentials are fitted to MD simulation data, as MD simulations also sample from non-equilibrium structure configurations. We compare the distribution  of forces and stresses in data sampled from random structure relaxations versus data sampled from MD simulations. In the experiments in section \ref{ml}, models were trained on all stoichiometries except for Li\textsubscript{15}Si\textsubscript{4}, which we held out for out-of-domain generalization experiment. The random structure relaxations have higher stresses than the MD simulations. The forces for the random structure relaxations, however, are smaller than those in the MD simulations. We enforced a moderately large minimum separation between atoms for random structure generation to ensure that none of the random structure relaxations diverged, and moved atoms in the direction of the forces during the relaxation. This led to a narrower distribution of forces in the random structure relaxation dataset.

\subsection{Machine learning random structure relaxations}\label{ml}
As a usage example of this dataset, we trained graph neural network (GNN) models to simulate structure relaxations. The bottleneck in ab initio random structure search is the DFT calculation of forces and stresses at each ionic relaxation step, which is computationally expensive. We replace the DFT calculation of forces and stresses with a machine learned model to speed up crystal structure search. We train our GNN model on the samples from relaxation trajectory with intermediate structures as input and DFT-calculated per-atom forces and unit cell stresses as output. Then, given a random structure, we can relax the structure using the force and stress outputs from the GNN using the same ionic relaxation mechanism as implemented in VASP. 

We used these models in conjunction with AIRSS to demonstrate an accelerated search for low-energy crystal structures. Ideally, given a random initial structure, the GNN model's relaxation should reproduce the same relaxation trajectory as DFT and yield the same relaxed structure. First, to evaluate how accurately our GNNs compute forces and stresses, we compute the force and stress mean absolute error from the test set. We compute the errors on individual structures in the test set, sampled as described in the Methods section. Second, we evaluate how well our GNN model reproduces the same final relaxed structures as DFT given the same random initial structure. At test time, only the initial structure is given to the GNN, and it does not have access to the DFT relaxation trajectory. Given a number of test structures, we report how many of the initial structures in the test set relax to the same structure as DFT. In some cases, the GNN relaxation trajectory initially follows the DFT relaxation trajectory, but starts to deviate during the relaxation. To capture and quantify this behavior, we also match each structure at the same step of the DFT and GNN relaxation trajectory, and record the percentage of the structures that match. We report the average percentage of the relaxation trajectory GNNs match DFT in the test set. 

To compute the metrics for matching trajectories, we use all the ionic relaxation steps until the maximum force in the crystal reaches 0.1 eV/\AA. The cutoff of 0.1 eV/\AA was imposed since our goal is to quickly screen random structures to find the few lowest energy structures, and $49.5\%$ of the relaxation trajectories match the converged final structure by the time the maximum force acting on the atoms reaches 0.1 eV/ \AA. Structure match is computed with the StructureMatcher module of pymatgen. As crystal structure descriptors such as space groups and lattice parameters do not uniquely determine the structure, StructureMatcher provides a more direct comparison by mapping one crystal structure's lattice and atoms onto another structure and quantifying  the deviation\footnote{see https://pymatgen.org/pymatgen.analysis.structure\_matcher.html}.

Next, we perform a series of experiments to explore how GNN models can be improved for crystal structure search and quantify how generalizable these models are.
\begin{itemize}
    \item GNN architectures for predicting unit cell stress (section \ref{architecture})
    \item Effects of adding Gaussian noise (section \ref{noise})
    \item Ablation experiments (section \ref{ablation})
    \begin{itemize}
        \item Effects of training on MD data
        \item Effects of training on energies
    \end{itemize}
    \item Out-of-domain generalizations (section \ref{generalization})
    \begin{itemize}
        \item Generalization across different stoichiometries
        \item Generalization across different DFT parameters
    \end{itemize}
    \item Application to crystal structure search (section \ref{css})
\end{itemize}

In section \ref{architecture} and \ref{noise}, we explore how we can improve the accuracy of stress calculations by modifying the GNN architecture and adding Gaussian noise during training. Then we take a deeper dive into what contributes the most to the GNN's superior performance in crystal structure search. In section \ref{ablation}, we quantify the effects of training on random structure data compared to MD data, and training directly on forces and stresses compared to energies. Finally, we examine how well our models generalize to systems with different stoichiometries as well as to data generated with different DFT parameters in section \ref{generalization}.

\subsubsection{Our GNN architecture is effective in simulating relaxation trajectories}\label{architecture}

We investigated three different architectures:

{\bf Stress tensor outputs as global output features.} The 6 independent components of the stress tensor($\sigma_{xx}, \sigma_{yy}, \sigma_{zz}, \sigma_{xy}, \sigma_{yz}, \sigma_{zx}$) are predicted from the graph. The node and edge states are aggregated with set2set~\cite{set2set} and concatenated with the global state, and passed through two MLP layers of sizes 64 and 32. The final predictions are produced by an output layer without activation. For this model, the distance from each atom to (100), (010) and (001) planes were concatenated to the node feature inputs.

{\bf Stress tensor outputs from fictitious atoms on cell corners.} We put fictitious non-interacting atoms on the corners of the cells, and predict the stress tensor components from the fictitious atoms. Each fictitious atom outputs three stress tensor components - e.g. the fictitious atom on the x-axis outputs $\sigma_{xx}, \sigma_{yx}, \sigma_{zx}$.

{\bf Stress tensor outputs from fictitious atoms on cell planes.} We put fictitious non-interacting atoms on (100), (010) and (001) planes, and use distances from the planes to each atom for computing edge features.

As a baseline, we use the second nearest-neighbor modified embedded atom method (2NN MEAM) from \cite{meam}. 

Table~\ref{stress} shows that stress predictions from fictitious atoms on cell planes and as global output perform similarly on force and stress MAE, but fictitious atoms on cell corners performs poorly on stress MAE. These two models outperform MEAM in both force and stress predictions by an order of magnitude. Moreover, fictitious atoms on cell planes is the best at predicting the relaxation trajectory, with 74\% improvement over the MEAM baseline. We used fictitious atoms on cell planes for all following experiments. 

\subsubsection{Gaussian noise improves both force predictions and relaxation trajectories}\label{noise}

We investigate the effects of adding Gaussian noise to atom positions and lattice vectors. In computer vision, there has been evidence of Gaussian noise increasing out-of-domain accuracy at the cost of in-domain accuracy \cite{gilmer2019adversarial}. The improvement in out-of-domain accuracy has been attributed to reduced sensitivity to high frequency noise~\cite{lopes_improving_2019, yin_fourier_2019}.  Using Gaussian noise during training~\cite{sanchez-gonzalez_learning_2020} and training on accuracy of configurations few steps in the future~\cite{ummenhofer_lagrangian_2019} have also been shown to improve roll-out accuracy of fluid simulations. In ~\cite{sanchez-gonzalez_learning_2020}, adding small perturbations during training was observed to make the simulations more robust over longer trajectories at the cost of accuracy at each time step.

As a baseline, we use the second nearest-neighbor modified embedded atom method (2NN MEAM) from \cite{meam}. All GNN models outperform MEAM by an order of magnitude difference in both force and stress MAE. In contrast to \cite{sanchez-gonzalez_learning_2020} where Gaussian noise helped simulate longer trajectories at the cost of force errors at each step (one-step MSE in their paper), in our system adding Gaussian noise leads to more accurate force and stress outputs and better relaxation trajectories. In table \ref{augment}, all metrics improve with Gaussian noise of standard deviation 0.03 \AA. The advantages of Gaussian noise decrease with larger amounts of noise. Recent work on adversarial examples in physical systems~\cite{cubuk2020adversarial} showed that models approximating physical systems move in their own adversarial directions, which might explain why adding Gaussian noise was advantageous in this work.

\subsubsection{Ablation experiments: Molecular dynamics cannot predict random structure relaxations}\label{ablation}
Compared to the MEAM baseline, our approach uses a more expressive model family (GNN) and a richer dataset. In order to determine which factors contribute most to our model's superior performance, we perform two ablation experiments. 

{\bf Baseline 1. } We train the same GNN architecture on a separate dataset of structures from a molecular dynamics trajectory. If the model trained on the MD dataset performs poorly on structure search, this suggests that training on our random structure relaxation dataset is necessary for structure search.

From our MD dataset, we use StructureMatcher to discard matching structures, as was done for relaxation trajectories of random structure relaxations data. We end up with 17134(LiSi), 16088(Li\textsubscript{13}Si\textsubscript{4}), 9613(Li\textsubscript{2}Si), 13079(Li\textsubscript{7}Si\textsubscript{2}) and 12749(Li\textsubscript{15}Si\textsubscript{4}) structures for each stoichiometry.

{\bf Baseline 2. } We train the GNN on the same dataset of random structure relaxations, but train on energies instead of forces and stresses. Comparing to the model trained on energies will tell us how much improvement was gained from our architecture that trains directly on forces and stresses.

The model trained on random structure energies achieve a MAE of 13.5 meV/atom. This is higher than other machine learning potentials on the Li-Si systems, such as \cite{artrith_constructing_2018, onat_implanted_2018, xu_deep-learning_2020}. As random structure data is quite different from what \cite{artrith_constructing_2018, onat_implanted_2018, xu_deep-learning_2020} are trained on, we trained and tested the same model on our MD data. The test MAE for MD energies is 3.8 eV/atom, similar to what's reported in these works. 

The results are in table~\ref{low}. Our architecture trained on random structure data shows an order of magnitude improvement against both baselines. The MD baseline suggests that the distribution of data generated by MD is quite different and training on random structures is crucial for crystal structure search.

\subsubsection{Data augmentation helps out-of-domain generalizations}\label{generalization}
\textbf{Generalization across different stoichiometries.}
We trained the models on four stoichiometries in the Li-Si system - LiSi, Li\textsubscript{2}Si, Li\textsubscript{7}Si\textsubscript{2}, Li\textsubscript{13}Si\textsubscript{4}. We tested the model on 4199 random structure relaxations of Li\textsubscript{15}Si\textsubscript{4} to see if it recovers the experimentally reported unit cell of Li\textsubscript{15}Si\textsubscript{4}. The experimental structure of Li\textsubscript{15}Si\textsubscript{4} has very different lattice parameters and symmetries from other experimental structures of the Li-Si system, as can be seen in table \ref{lisi_data} and in the visualization in the supporting information. 

We ensure that there are no structures in the training set that match the experimental structure of Li\textsubscript{15}Si\textsubscript{4} by using a training set with a smaller number of atoms and different stoichiometries. The experimental structure of Li\textsubscript{15}Si\textsubscript{4} has 38 atoms in the unit cell, which is larger than any of the unit cells used in training. This also makes it a more challenging task, as the number of possible configurations scales exponentially with the number of atoms in the cell. Moreover, by using different a different stoichiometry, the ratio of Li to Si within the same cell differs from the training set to the test set.

Again, we compare it to the 2NN MEAM baseline. Out of the 4199 random structure relaxations using DFT, three of the structures relaxed into the same experimental structure reported in \cite{dupke_structural_2012}, and the structures relaxed to the force cutoff of 0.1 eV/\AA{}  match the final relaxed structures. The structures relaxed with DFT were checked with pymatgen's StructureMatcher with default parameters to identify identical structures, which checks if two structures mapped onto another are the same up to a small displacement in atom positions and lattices. Though the space groups of the relaxed structures did not exactly match the experimental structure, the structures were close enough to be matched by StructureMatcher, with up to 0.1\% difference in volume.

All GNN models reproduced the experimental structure trajectories, and MEAM did not find the experimental structure, as shown in table~\ref{ood_stoichiometry}. This is surprising since Li\textsubscript{15}Si\textsubscript{4} was included in the data used to fit MEAM. It suggests that for random structure relaxations, the behavior of the force field model at high-stress regions may relax the structure into the wrong basin in the energy landscape and training on this region may be crucial for random structure search.

\begin{figure}
  \centering
  \includegraphics[width=\linewidth]{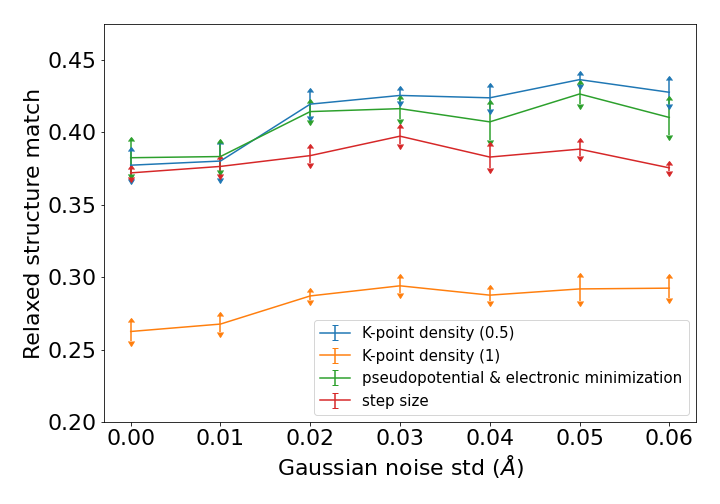}
  \caption{\textbf{Out-of-domain generalization on Li\textsubscript{15}Si\textsubscript{4} structure relaxations generated with different DFT parameters.} K-point density units are \AA\textsuperscript{-1}.
  We train models on data generated with default DFT parameters, and test how well they generalize on data generated with different DFT parameters. The difference in data distribution generated with different DFT parameters and default DFT parameters is plotted in figure~\ref{OOD_data}. A decrease in model performance from table~\ref{ood_stoichiometry} may be expected as the data comes from different distributions. We see a 30\% decrease in relaxed structure match from table~\ref{ood_stoichiometry} for data generated with K-point density 1 \AA\textsuperscript{-1}, which deviates substantially from the training distribution. However, Gaussian noise helps generalization in all cases, especially to data with K-point density 0.5 \AA\textsuperscript{-1} and different pseudopotential at large amounts of Gaussian noise. Error bars are computed from three different random seeds, dividing the standard deviation by the square root of number of measurements. }\label{ood-vasp}
\end{figure}

\textbf{Generalization across different DFT parameters.}
In many datasets of materials, different DFT parameters may be used for the same material. How robust are the models trained on our dataset to changes in DFT patameters? We test the models on our out-of-domain dataset generated with different parameters used for DFT. The initial structures relaxed with different DFT parameters are a subset of Li\textsubscript{15}Si\textsubscript{4} structures above, so this is an out-of-domain generalization in both DFT parameters and stoichiometry. Our results for different DFT parameters are plotted in figure~\ref{ood-vasp}. 

We see that Gaussian noise helps generalization in all cases. As the data generated with k-point density of 1 \AA\textsuperscript{-1} deviates the most from the settings used in training data, the model tested on k-point density of 1 \AA\textsuperscript{-1} performs the worst as expected. Curiously, while the data generated with k-point density of 0.5 \AA\textsuperscript{-1} has final energy MAE an order of magnitude larger than data generated with different pseudopotential and step size, generalization to k-point density of 0.5 \AA\textsuperscript{-1} is better. 

\subsubsection{Accelerating crystal structure search with machine learning}\label{css}

VASP simulations of Li\textsubscript{15}Si\textsubscript{4} structures took 38 hours on average on 8-core N1 machines on Google Cloud. GNN simulations take less than a minute on a laptop (MacBook Pro (with 2.8GHz quad-core Intel Core i7, 2017)), and is still able to find the experimental structure of Li\textsubscript{15}Si\textsubscript{4}. By making the calculations more efficient, we are able to explore the applicability of AIRSS to large scale unit cells. Our approach to speeding up AIRSS can find optimal structures with 38- atom Li\textsubscript{15}Si\textsubscript{4} with 2000 trials, but we were unable to find the experimental structure of 152-atom Li\textsubscript{12}Si\textsubscript{7}. We generated 160,000 random structures of Li\textsubscript{12}Si\textsubscript{7} with 152 atoms per unit cell and relaxed the structures using our model with Gaussian noise 0.03\AA. As the experimental structure of Li\textsubscript{12}Si\textsubscript{7} belongs to space group Pnma with 8 symmetry elements, we generated half of the random structures (80,000) with 8 symmetry elements to increase the chance of finding the experimental structure. A similar experiment was also done in \cite{deringer_data-driven_2018}, where the authors did 100,000 random relaxations and failed to find fibrous P with 21 symmetry-independent atoms. Though our search for Li\textsubscript{12}Si\textsubscript{7} with 8 symmetry elements have 22 symmetry-independent atoms similar to fibrous P, having two atom types complicates our search even further. Both results indicate that though structural relaxations can be made more efficient with machine learning, random structure search for large unit cells is still limited by combinatorial explosion, and may require additional constraints such as the ones used by \cite{deringer_hierarchically_2020}. 

\section*{Discussion}
 Crystal structure search is a fundamental piece of materials design, but research interest from machine learning communities has been relatively small. We build a novel dataset of DFT random structure relaxations, and show that graph network models trained on this data has significant advantages for crystal structure search. We hope to encourage the community of researchers in machine learning for materials science to study the problem of crystal structure search, and broaden the understanding of energy landscape of crystalline materials. We leave the expansion of this work to the ternary Li-Si-O system for future work. Moreover, we show that using Gaussian noise in training force field models improve accuracy in force predictions, simulating trajectories and out-of-domain generalization. It would be interesting to draw on the literature on adversarial defense and explore how it can be used to improve physical models. 
 
\section*{Methods}

\subsection*{DFT calculations}

We use AIRSS~\cite{airss2006,airss2011} to generate random unit cells. Additional constraints on the cell volume, minimum separation between atoms and symmetry are imposed on the cell as described in~\cite{airss2006,airss2011}. We generate random structures with up to 6 symmetry operations. We vary the number of formula units in the cell up to 38 atoms in the cell. We estimate the volume per element from the experimental data of Li-Si structures in the Materials Project, and sample the target volume of AIRSS structure outputs uniformly at random between 80\% to 150\% of the estimated volume (unless indicated otherwise). We set the minimum separation between atoms to be 2.28 \AA, as small values of atomic separation can cause forces to diverge. 

The Vienna Ab initio Simulation Package (VASP)\cite{kresse_software_1993} is used to relax random structures via the MedeA software environment~\cite{medea}. We use Kohn-Sham density functional theory with the projector augmented-wave method\cite{blochl_projector_1994} as implemented in VASP,  with self-consistent, periodic DFT. We use the steepest descent algorithm (IBRION=3, SMASS=2, POTIM=0.4) to relax both the ions and the unit cell(ISIF=3). The electronic minimization algorithm was set to the default (ALGO=Normal, blocked Davidson iteration scheme). The energy convergence threshold is set to 10\textsuperscript{-6} eV and force convergence to 10\textsuperscript{-2} eV/\AA. We use a plane wave basis set with the kinetic energy cutoff of 520 eV, and we sample the Brillouin zone with a $\Gamma$-centered k-point mesh with spacing 0.25 \AA\textsuperscript{-1}. We use a generalized gradient approximation Perdew-Burke-Ernzerhof (PBE)\cite{perdew_generalized_1996} functional to treat the exchange-correlation energy, using the  Li\_sv and Si pseudopotentials as implemented in VASP.

Molecular dynamics simulations are also carried out using VASP. The simulations were run at a constant temperature of 800K with 2 femtosecond time steps, with a single k-point at Gamma for at least 47 picoseconds. All other DFT parameters remain the same as the ones used in the random structure relaxation dataset.

\subsection*{Data selection}
We train all the ML models in this section on random structures in the Li-Si system, selecting batches of 5000 random structure relaxations of each stoichiometry. The total number of converged random structure relaxations for each stoichiometry were 4680(LiSi), 4365(Li\textsubscript{13}Si\textsubscript{4}), 4596(Li\textsubscript{2}Si), 4515(Li\textsubscript{7}Si\textsubscript{2}) and 4199(Li\textsubscript{15}Si\textsubscript{4}). We use only the converged trajectories in our experiments.  We hold out Li\textsubscript{15}Si\textsubscript{4} for out-of-domain stoichiometry generalization experiments, and train on the rest of the Li-Si stoichiometries, which are LiSi, Li\textsubscript{13}Si\textsubscript{4}, Li\textsubscript{2}Si and Li\textsubscript{7}Si\textsubscript{2}. The reference structures we used for the computations can be found in table \ref{lisi_data}.

We construct our training set by sampling structures from the relaxation trajectory to avoid samples very similar in structure and energy. In this work we used pymatgen's~\cite{pymatgen} StructureMatcher with (stol = 0.05) to define matching structures.\footnote{See https:\/\/pymatgen.org\/pymatgen.analysis.structure\_matcher.html } We sample at most one structure out of every five ionic steps. We only sample structures that do not match the previous structure in the trajectory. After sampling, we used 80\% of the trajectories for training and 10\% each for validation and test. We use these samples for evaluating mean absolute errors of forces and stresses, which allows us to avoid biases from having many similar structures in the test set. 

\subsection*{Model training details}
Our model is based on MEGNet\cite{megnet} with the following modifications. Each node is initialized to a one-hot vector corresponding to the atomic number of the element, and a 16-dimensional embedding layer is applied. Edges are constructed for every pair of atoms within 4\AA, and each edge is initialized to the displacement vector between the atoms, distance, and distance squared ($(\vec{r_{ij}}, |\vec{r_{ij}}|, |\vec{r_{ij}}|^2)$, where $\vec{r_{ij}}=\vec{r_{i}}-\vec{r_{j}}$). Global states are initialized to 0. Each state is passed through the 7 message-passing layers implemented in MEGNet, consisting of two MLP layers of size 64 and state updates. To get forces on each atom, we concatenate node features(with two MLP layers of size 64 and 32) with global features and follow with a output layer without any activation function. For the stress tensor predictions, we investigated three architectures and chose the one with best performance described in section~\ref{architecture}.

The force and stress inputs were scaled to have mean 0 and standard deviation 1. We used the sum of mean squared errors of the forces and stresses for our loss function. We augmented our data by rotating 50\% of the structures at each epoch by 90 degrees. For 30\% of the structures, we perturbed the atom positions and lattice vectors with Gaussian noise. The models were trained with ADAM optimizer with initial learning rate $10^{-3}$. We found that the amount of Gaussian noise affects the optimal learning rate schedule, and models with larger amounts of noise needs slower learning rate decay. We halved learning rates every 15 to 35 epochs depending on the amount of Gaussian noise, with details in the supporting information. Using the force and stress outputs, we simulate relaxations using the velocity Verlet equations as implemented in VASP, details can be found in the supporting information. 

To construct the model trained on energies in section \ref{ablation}, we apply the same modification to our model trained on forces and stresses. After the message-passing layers, the node and edge states are aggregated with set2set~\cite{set2set} and concatenated with the global state, and passed through two MLP layers of sizes 64 and 32. The final predictions are produced by an output layer without activation. The energy inputs were also scaled to have mean 0 and standard deviation 1, and sum of mean squared errors was used as loss function. We used rotation augmentation for training the energy model. 
 
\section*{Data Availability}
Our dataset is available on a Google Cloud storage bucket at gs://gresearch/crystal-relaxations/. Instructions on downloading the data as well as Colab notebooks with working examples are available in the supporting information.

\section*{Author Contributions}
G.C. and L.Y. performed density functional theory calculations and prepared the dataset. G.C., K.M. and L.Y. performed machine learning simulations and analyzed the results. E.D.C. and L.Y. conceived and supervised the project. G.C., L.Y. and E.D.C. wrote the manuscript and revised the manuscript with K.M. and E.J.R. 

\section*{Competing Interests}
The Authors declare no competing interests.
 
 \section*{Acknowledgements}
 We would like to thank Michael Brenner for helpful discussions, and Samuel Schoenholz for giving feedback on the paper.  E. J. R. acknowledges support from Stanford's Storage X program.

\begin{landscape}
\begin{table}
  \caption{{\bf Summary of our dataset.} We relax batches of random structures with numbers indicated in {\it initial structures}. Some jobs may fail due to cloud errors or due to exceeding the time limit, and the number of jobs that successfully finished are in {\it jobs completed}. {\it Converged without error} is the  number of jobs that converged to the final structure below energy and force convergence criteria specified in the main text. {\it No ionic convergence}  jobs did not finish converging, but the relaxation trajectory still contains information on the energy landscape. {\it other VASP errors} are jobs that failed due to various errors in VASP relaxations. \\
  {\it Volume per formula unit}: 80\%\~150\% of the estimated volume from experimental data of Li-Si structures in the Materials Project, unless otherwise indicated. 200\%=200\% of the estimated volume. \\
  All calculations were done on Google cloud N1 machines, with the number of cores specified in {\it cores}.
  }\label{data_summary}
\centering
\begin{tiny}
\resizebox{9in}{!}{%
\begin{tabular}{|p{1cm}|p{1.5cm}|p{1.5cm}|p{0.7cm}|p{0.7cm}|p{0.7cm}|p{0.7cm}|p{0.7cm}|p{0.7cm}|p{0.7cm}|p{1cm}|p{0.7cm}|p{1cm}|p{0.7cm}|p{1cm}|}
\toprule
formula & {batch} & folder & {initial structures} & {volume per formula unit} & {formula units} & {jobs completed} & {converged without error} & {no ionic convergence} & {other VASP errors} & {mean number of steps to converge} & {maximum number of ionic steps} & {total computation time (hours)} & {cores} & {total core-hours} \\ \midrule
LiSi & 2 & Li1Si1\_02/ & 5000 & {320 \AA\textsubscript{3}} & 8 & 4999 & 4853 & 93 & 53 & 548.03 & 2000 & 5.18E+04 & 8 & 4.14E+05 \\ \hline
LiSi & 3 & Li1Si1\_03/ & 3000 & {400 \AA\textsubscript{3}} & 8 & 2998 & 2875 & 77 & 46 & 581.66 & 2000 & 6.38E+04 & 8 & 5.11E+05 \\ \hline
LiSi & 4 & Li1Si1\_04/ & 2000 & {560 \AA\textsubscript{3}} & 8 & 2000 & 1912 & 60 & 28 & 620.90 & 2000 & 2.89E+04 & 8 & 2.31E+05 \\ \hline
LiSi & 5 & Li1Si1\_05/ & 5000 & - & 1-16 & 4992 & 4680 & 260 & 52 & 558.11 & 2000 & 9.47E+04 & 8 & 7.57E+05 \\ \hline
Li2Si & 2 & Li2Si1\_02/ & 5000 & - & 1-12 & 4992 & 4596 & 343 & 53 & 614.15 & 2000 & 1.15E+05 & 8 & 9.21E+05 \\ \hline
Li2Si & 3 & Li2Si1\_03/ & 5000 & - & 2-4 & 5000 & 4898 & 5 & 97 & 317.56 & 2000 & 1.69E+04 & 8 & 1.35E+05 \\ \hline
Li7Si2 & 3 & Li7Si2\_03/ & 5000 & - & 1-4 & 4983 & 4515 & 412 & 56 & 685.26 & 2000 & 1.31E+05 & 8 & 1.05E+06 \\ \hline
Li7Si2 & 4 & Li7Si2\_04/ & 5000 & - & 4 & 4991 & 3980 & 975 & 36 & 960.60 & 2000 & 1.56E+05 & 16 & 2.50E+06 \\ \hline
Li7Si2 & 5 & Li7Si2\_05/ & 5000 & - & 4 & 4993 & 3972 & 967 & 54 & 950.25 & 2000 & 1.51E+05 & 16 & 2.42E+06 \\ \hline
Li13Si4 & 2 & Li13Si4\_02/ & 10000 & {200\%} & 2 & 9639 & 7572 & 1956 & 111 & 1,016.14 & 2000 & 7.06E+05 & 8 & 5.65E+06 \\ \hline
Li13Si4 & 3 & Li13Si4\_03/ & 5000 & - & 1-2 & 4981 & 4365 & 558 & 58 & 763.60 & 2000 & 1.61E+05 & 8 & 1.29E+06 \\ \hline
Li15Si4 & 2 & Li15Si4\_02/ & 5000 & - & 1-2 & 4916 & 4299 & 548 & 69 & 812.51 & 2000 & 1.90E+05 & 8 & 1.52E+06 \\ \hline
Li1O1 & 1 & Li1O1\_01/ & 5000 & - & 1-16 & 4221 & 2086 & 1798 & 337 & 399.45 & 2000 & 3.06E+05 & 8 & 2.45E+06 \\ \hline
Li2O & 1 & Li2O1\_01/ & 5000 & - & 1-12 & 4906 & 4590 & 141 & 175 & 474.18 & 2000 & 9.87E+04 & 8 & 7.90E+05 \\ \hline
SiO2 & 3 & SiO2\_03/ & 10000 & - & 1-12 & 9160 & 138 & 1091 & 7931 & 245.10 & 2000 & 1.79E+05 & 8 & 1.43E+06 \\ \hline
SiO2 & 4 & SiO2\_04/ & 10000 & - & 1-12 & 9620 & 152 & 1504 & 7964 & 308.62 & 2000 & 2.48E+05 & 8 & 1.99E+06 \\ \hline
SiO2 & 5 & SiO2\_05/ & 1000 & - & 2 & 992 & 443 & 535 & 14 & 1,033.90 & 2000 & 4.13E+04 & 32 & 1.32E+06 \\ \hline
SiO2 & 6 & SiO2\_06/ & 6000 & - & 2-4 & 5995 & 5238 & 597 & 160 & 1,019.71 & 3000 & 6.68E+04 & 32 & 2.14E+06 \\ \hline
SiO2 & 7 & SiO2\_07/ & 2000 & - & 6 & 1996 & 1384 & 600 & 12 & 1,492.28 & 3000 & 6.56E+04 & 32 & 2.10E+06 \\ \hline
SiO2 & 8 & SiO2\_08/ & 2000 & - & 8 & 1922 & 1102 & 792 & 28 & 1,619.18 & 3000 & 1.18E+05 & 32 & 3.78E+06 \\ \hline
SiO2 & 9 & SiO2\_09/ & 2000 & - & 12 & 1733 & 811 & 915 & 7 & 1,884.37 & 3000 & 1.92E+05 & 32 & 6.14E+06 \\ \hline
Li4SiO4 & 1 & Li4Si1O4\_01/ & 5000 & - & 2 & 4887 & 353 & 3276 & 1258 & 350.69 & 2000 & 2.64E+05 & 32 & 8.44E+06 \\ \hline
Li2Si2O5 & 1 & Li2Si2O5\_01/ & 5000 & - & 2 & 4979 & 3452 & 1431 & 96 & 1,549.92 & 3000 & 1.49E+05 & 32 & 4.75E+06 \\ \hline
Li15Si4 & {Li pseudopotential, conjugate algorithm} & Li15Si4\_02\_fa/ & 800 & - & 1-2 & 798 & 675 & 114 & 9 & 787.60 & 2000 & 1.52E+04 & 96 & 1.46E+06 \\ \hline
Li15Si4 & {k-point density(0.5)} & Li15Si4\_02\_k/ & 800 & - & 1-2 & 799 & 678 & 112 & 9 & 820.59 & 2000 & 4.07E+03 & 96 & 3.91E+05 \\ \hline
Li15Si4 & {k-point density(1)} & Li15Si4\_02\_k2/ & 800 & - & 1-2 & 800 & 690 & 100 & 10 & 853.69 & 2000 & 1.31E+03 & 96 & 1.25E+05 \\ \hline
Li15Si4 & {step size (0.8)} & Li15Si4\_02\_p/ & 800 & - & 1-2 & 800 & 781 & 8 & 11 & 525.45 & 2000 & 9.04E+03 & 96 & 8.68E+05 \\ \hline
TOTAL & {} &  & 116200 &  & {} & {} & {} & {} & {} & {} & {} & {} & {} & 5.56E+07 \\ \bottomrule

\end{tabular}%
}
\end{tiny}
\end{table}
\end{landscape}

\begin{table}
  \caption{{\bf Experimentally determined crystal structures in the Li-Si system.} We selected 5 Li-Si systems in the Materials Project that have corresponding experimentally determined structures in the ICSD: LiSi, Li\textsubscript{2}Si, Li\textsubscript{7}Si\textsubscript{2}, Li\textsubscript{13}Si\textsubscript{4} and Li\textsubscript{15}Si\textsubscript{4}. For MD simulations, we created supercells of each experimental unit cell to ensure that the cells are sufficiently large to capture the dynamics.}
  \label{lisi_data}
  \centering
\begin{tabular}{cccccc}
\toprule 
Formula & Space group & Lattice parameters & Reference & Materials & \# atoms in \\
 &  & a, b, c (\AA) &  & Project ID & MD supercell \\
 \midrule
LiSi & I41/a & 9.35, 9.35, 5.74 & \cite{evers_lisi_1997} & mp-795 & 64 \\
Li\textsubscript{7}Si\textsubscript{2} & Pbam & 7.99, 15.21, 4.43 & \cite{schafer_kristallstruktur_1965} & mp-27930 & 72 \\
Li\textsubscript{2}Si & C2/m & 7.70, 4.41, 6.56 & \cite{axel_kristallstruktur_1965} & mp-27705 & 72 \\
Li\textsubscript{13}Si\textsubscript{4} & Pbam & 7.95, 15.12, 4.47 & \cite{zeilinger_revision_2013} & mp-672287 & 68 \\
Li\textsubscript{15}Si\textsubscript{4} & I-43d & 10.63, 10.63, 10.63 & \cite{zeilinger_stabilizing_2013} & mp-569849 & 76 \\ 
\bottomrule
\end{tabular}%
\end{table}

\begin{table}
  \caption{{\bf Distribution of the random relaxations and molecular dynamics datasets}. Force and stress refers to the statistics of the components of force vectors and stress tensors. Max \(|\)force\(|\)  refers to the maximum force magnitude in each structure. Max stress component is the maximum value of the stress tensor components ($\sigma_{xx}, \sigma_{yy}, \sigma_{zz}, \sigma_{xy}, \sigma_{yz}, \sigma_{zx}$) in each structure. Many interatomic potentials are fitted to MD simulation data, as MD simulations also sample from non-equilibrium structure configurations. Comparing the data from the two sampling methods shows that the distributions are different, with random relaxations having significantly higher stresses than MD simulations.}
  \label{data_details}
  \centering
  \begin{tabular}{lllll}
    \toprule             
    Random relaxations 
     & mean    & standard deviation    & min   & max\\
    \midrule
    Force (eV/\AA)  &   0.000 & 0.208 & -4.797 & 4.797   \\
    Stress  (kBar) & 0.733 & 12.318 & -192.012 & 439.783   \\
    max \(|\)force\(|\) (eV/\AA) & 0.453 & 0.442 & 0.000 & 4.797   \\
    max stress component (kBar) & 8.662 & 23.555 & 0.000 & 439.783 \\
    \midrule
    Molecular dynamics 
     & mean    & standard deviation    & min   & max\\
     \midrule
    Force (eV/\AA)  & 0.000 & 0.523 & -5.347 & 6.355 \\
    Stress  (kBar) & 5.858 & 12.326 & -39.580 & 63.525 \\
    max \(|\)force\(|\) (eV/\AA)  & 0.424 & 0.412 & -2.598 & 6.355 \\
    max stress component (kBar) & 29.492 & 6.267 & 6.115 & 63.525 \\
    \midrule
    \bottomrule
  \end{tabular}
\end{table}

\begin{table*}
  \caption{{\bf Methods for predicting cell stress.} 
  We investigated three different architectures for predicting stresses on unit cells.\\ \textbf{Corner}: We put fictitious non-interacting atoms on the corners of the cells, and predict the stress tensor components from the fictitious atoms. Each fictitious atom outputs three stress tensor components - e.g. the fictitious atom on the x-axis outputs $\sigma_{xx}, \sigma_{yx}, \sigma_{zx}$.  \textbf{Plane}: We put fictitious non-interacting atoms on (100), (010) and (001) planes, and use distances from the planes to each atom for computing edge features. \textbf{Global output}: The 6 independent components of the stress tensor($\sigma_{xx}, \sigma_{yy}, \sigma_{zz}, \sigma_{xy}, \sigma_{yz}, \sigma_{zx}$) are predicted from the graph. \textbf{MEAM}: second nearest-neighbor modified embedded atom method (2NN MEAM) from \cite{meam}. \\
  Quantities in parentheses are the standard deviations from different random seeds. \textbf{*}: Fraction of relaxed final structures that match VASP \textbf{**}: Fraction of matching structures in each trajectory.}
  \label{stress}
  \centering
  \begin{tabular}{lllll}
    \toprule             
    Method    & Force MAE    & Stress MAE    & Relaxed structure& Matches in\\
    &&&match(\%)*&trajectory(\%)**\\
    \midrule
    corner    & 0.026 (0.0009)	& 2.32 (0.006) & 15 (0.7) 	& 33 (0.3) \\ 
     planes    & \textbf{0.025 (0.0005)}	& \textbf{0.75 (0.032)}	& \textbf{40 (0.7)}	& \textbf{66 (0.6)} \\ 
    global output               & 0.032 (0.0014) & 0.79 (0.053)  & 38 (1.4)	& 62 (0.7) \\ 
    MEAM               & 0.55 & 9.11  &  23 & 39 \\
    \bottomrule
  \end{tabular}
\end{table*}

\begin{table*}[t]
  \caption{{\bf Gaussian noise improves both force predictions and relaxation trajectories.} The units for Gaussian noise standard deviation, force and stress are \AA, eV/\AA{} and kBar. We averaged the results from three random seeds. Quantities in parentheses are the standard deviations from different random seeds. The last row uses MEAM from \cite{cui_second_2012}.
  *: Fraction of relaxed final structures that match VASP **: Fraction of matching structures in each trajectory \\
  All GNN models outperform MEAM by an order of magnitude difference in both force and stress MAE. Adding Gaussian noise leads to more accurate force and stress outputs and better relaxation trajectories. The advantages of Gaussian noise decrease with larger amounts of noise. 
}
  \label{augment}
  \centering
  \begin{tabular}{lp{2.2cm}llp{2.5cm}p{2.5cm}} 
    \toprule             
    Model & Gaussian noise (std. dev)   & Force MAE    & Stress MAE    & Relaxed structure match(\%)* & Matches in trajectory(\%)** \\  
    \midrule
GNN & 0.0	& 0.025 (0.0005)	& 0.75 (0.032)	& 40 (0.7)	& 66 (0.6)\\
GNN & 0.01	& 0.025 (0.0002)	& 0.78 (0.011)	& 41 (1.3)	& 66 (1.2)\\
GNN & 0.02	& 0.025 (0.0008)	& \textbf{0.73} (0.042)	& 42 (0.2)	& 66 (0.5)\\
GNN & 0.03	& \textbf{0.024} (0.0007)	& \textbf{0.73} (0.023)	& \textbf{44} (0.4)	& \textbf{68} (0.5)\\
GNN & 0.04	& 0.025 (0.0009)	& 0.74 (0.027)	& \textbf{44} (1.3)	& \textbf{68} (1.3)\\
GNN & 0.05	& 0.025 (0.0009)	& 0.76 (0.042)	& \textbf{44} (1.4)	& 67 (1.7)\\
GNN & 0.06	& 0.025 (0.0006)	& 0.76 (0.027)	& 43 (1.3)	& 67 (1.4)\\
MEAM & -    & 0.55              & 9.11          &  18         & 39 \\
    \bottomrule
    \vspace{-0.6cm}
  \end{tabular}
\end{table*}

\begin{table}
  \caption{{\bf Ablation experiments for using random structure relaxations data and using our architecture for directly predicting forces and stresses.} Compared to the MEAM baseline, our approach uses a more expressive model family (GNN) and a richer dataset. In order to determine which factors contribute most to our model's superior performance, we perform two ablation experiments. \\
  Comparison with molecular dynamics data: We train the same GNN architecture on a separate dataset of structures from a molecular dynamics trajectory. \\
  Comparison with a model trained on energies: Comparing to the model trained on energies will tell us how much improvement was gained from our architecture that trains directly on forces and stresses. \\
  First row is the result from table~\ref{augment} without Gaussian noise augmentation. RR = random relaxations.
  }
  \label{low}
  \centering
  \begin{tabular}{p{0.2\linewidth}p{0.1\linewidth}p{0.1\linewidth}p{0.1\linewidth}p{0.1\linewidth}p{0.1\linewidth}} 
    \toprule             
    Model  & Trained on    &Training data & Test data& Force MAE     & Stress MAE \\
    \midrule
    GNN     &forces stresses &  RR & RR & \textbf{0.025} 	& \textbf{0.75}	\\ \hline
    \multirow{2}{*}{GNN(1)}     & forces   & MD        & MD            &  0.067        &  1.63 \\
    & stresses &  MD  & RR   &  0.541         &  29.7 \\ \hline
    GNN(2)    & energies        & RR     & RR         &  0.227         &  19.1	\\ \hline
    MEAM    & -                     &  -        & RR         & 0.55          &  9.11\\
    \bottomrule
    \vspace{-0.6cm}
  \end{tabular}
\end{table}

\begin{table*}
  \caption{{\bf Generalization of the model on Li\textsubscript{15}Si\textsubscript{4}}. We train models on four stoichiometries in the Li-Si system excluding Li\textsubscript{15}Si\textsubscript{4}, and test how well it generalizes to Li\textsubscript{15}Si\textsubscript{4}. Models trained with Gaussian noise data augmentation are able to match the the ground truth structure and trajectory as calculated by VASP with higher efficiency. All of the machine learned force fields (GNN's) are able to find the experimentally verified structure.  *: whether relaxations with the model found the experimental structure of Li\textsubscript{15}Si\textsubscript{4}}
  \label{ood_stoichiometry}
  \centering
  \vspace{.5cm}
  \begin{tabular}{lllll}
    \toprule             
    Model & Gaussian noise    & Relaxed structure& Matches in& Experimental*\\
    &(std. dev)&match (\%)&trajectory (\%)&\\
    \midrule
    GNN & 0.0	& 38 (0.8)	& 66 (0.8)	& yes\\
    GNN & 0.01	& 38 (1.5)	& 66 (1.4)	& yes\\
    GNN & 0.02	& 41 (0.8)	& 68 (0.8)	& yes\\
    GNN & 0.03	& \textbf{42} (1.2)	& \textbf{69} (0.9)	& yes\\
    GNN & 0.04	& \textbf{42} (1.5)	& \textbf{69} (1.1)	& yes\\
    GNN & 0.05	& \textbf{42} (1.8)	& \textbf{69} (1.5)	& yes\\
    GNN & 0.06	& 41 (1.5)	& 68 (0.9)	& yes\\
    MEAM& -     & 19          & 39          & no\\
    \bottomrule
  \end{tabular}
  \vspace{.5cm}
\end{table*}

\clearpage
\printbibliography

\end{document}


\onecolumn
\icmltitle{Supplementary Information for Crystal Structure Search with Random Relaxations Using Graph Networks}



\icmlsetsymbol{equal}{*}

\begin{icmlauthorlist}
\icmlauthor{Gowoon Cheon}{to}
\icmlauthor{Lusann Yang}{goo}
\icmlauthor{Kevin McCloskey}{goo}
\icmlauthor{Evan J. Reed}{to}
\icmlauthor{Ekin D. Cubuk}{goo}
\end{icmlauthorlist}

\icmlaffiliation{to}{Stanford University, Stanford, California, USA}
\icmlaffiliation{goo}{Google Research, Mountain View, California, USA}

\icmlcorrespondingauthor{Gowoon Cheon}{gcheon@stanford.edu}
\icmlcorrespondingauthor{Ekin D. Cubuk}{cubuk@google.com}

\icmlkeywords{Materials Science, Computational Physics, Graph Neural Networks, Crystal Structure}

\vskip 0.3in


\section{Data summary}
Summary of our random structure relaxations dataset are in table~\ref{ds}, including the out-of-distribution data generated with different DFT parameters. 

To generate this data, we use the Vienna Ab initio Simulation Package (VASP)\cite{kresse_software_1993} to perform DFT relaxations on the random unit cells.The random structures are relaxed using Kohn-Sham density functional theory with the projector augmented-wave method\cite{blochl_projector_1994} as implemented in VASP. Relaxations are performed with steepest descent algorithm (IBRION=3, SMASS=2, POTIM=0.4) to relax both the ions and the unit cell(ISIF=3). Electronic minimization algorithm was set to default (ALGO=Normal, blocked Davidson iteration scheme). The calculations are converged to energy convergence of 10\textsuperscript{-6} eV and force convergence of 10\textsuperscript{-2} eV/\AA  with self-consistent, periodic DFT. A plane wave basis set with the kinetic energy cutoff of 520 eV is used, and $\Gamma$-centered k-point mesh with spacing 0.25 \AA\textsuperscript{-1} is used to sample the Brillouin zone. Generalized gradient approximation Perdue-Burke-Ernzerhof (PBE)\cite{perdew_generalized_1996} functional is used to treat the exchange-correlation energy, with pseudopotentials Li\_sv and Si as implemented in VASP.

\section{Estimation of number of configurations in unit cells}
The number of configurations for placing N atoms in a cell of volume V, discretizing the cell by length $\delta$, is $\frac{1}{(V/\delta^3)} \frac{(V/\delta^3)!}{(V/\delta^3-N)!N!}$ \cite{oganov_structure_2019}. For N=10, V=10\AA\textsuperscript{3} and $\delta$=0.1\AA, this number is 2.7$\times$ 10\textsuperscript{25}

\section{Distribution of random structure relaxations data and MD data}
\begin{figure}[hbt!]
  \centering
  \includegraphics[width=6cm]{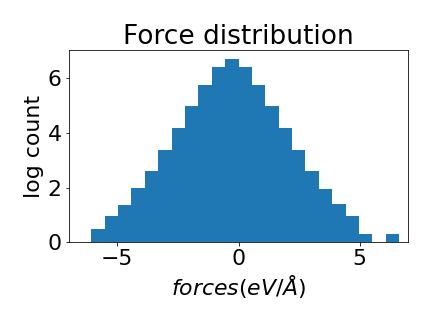}
  \includegraphics[width=6cm]{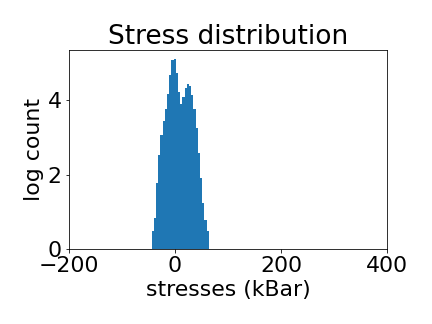}
  \caption{Force(left) and stress(right) distribution of MD data.}\label{md-dist}
\end{figure}

\begin{figure}[hbt!]
  \centering
  \includegraphics[width=6cm]{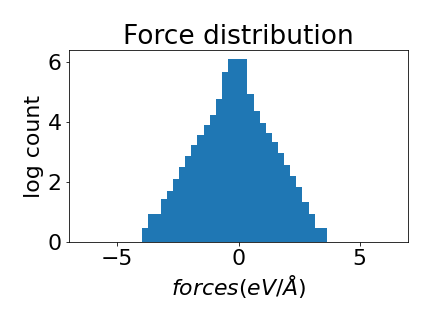}
  \includegraphics[width=6cm]{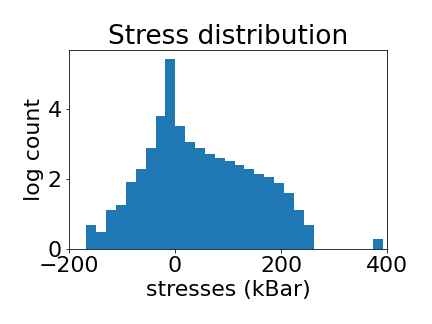}
  \caption{Force(left) and stress(right) distribution of random structure relaxations data.}\label{rss-dist}
\end{figure}

The distribution of forces and stresses for random structure relaxations data is summarized in table~\ref{data_details}. In the experiments in the main text, models were trained on all stoichiometries except for Li\textsubscript{15}Si\textsubscript{4}, which was held out for generalization experiments. The distributions of molecular dynamics and random structure relaxations are compared in in figure~\ref{md-dist} and figure~\ref{rss-dist}, which are also plotted excluding Li\textsubscript{15}Si\textsubscript{4}. We see that the random structures have higher stresses than MD data, which is expected from random structures. The forces for random structures, however, are smaller than MD data. We enforced a moderately large minimum separation between atoms for random structure generation to ensure that none of the random structures diverged during relaxations, and moved atoms in the direction of forces during the relaxations. This led to a narrower distribution of forces in the random structure dataset.

\begin{landscape}
\begin{table}
  \caption{{\bf Summary of our dataset.} We relax batches of random structures with numbers indicated in {\it initial structures}. Some jobs may fail due to cloud errors or due to exceeding the time limit, and the number of jobs that successfully finished are in {\it jobs completed}. {\it Converged without error} is the  number of jobs that converged to the final structure below energy and force convergence critea specified in the main text. {\it No ionic convergence}  jobs did not finish converging, but the relaxation trajectory still contains information on the energy landscape. {\it other VASP errors} are jobs that failed due to various errors in VASP relaxations. \\
  {\it Volume per formula unit}: 80\%\~150\% of the estimated volume from experimental data of Li-Si structures in the Materials Project, unless otherwise indicated. 200\%=200\% of the estimated volume. \\
  All calculations were with the number of cores specified in {\it cores}.
  }\label{ds}
\centering
\begin{tiny}
\resizebox{9in}{!}{%
\begin{tabular}{|p{1cm}|p{1.5cm}|p{1.5cm}|p{0.7cm}|p{0.7cm}|p{0.7cm}|p{0.7cm}|p{0.7cm}|p{0.7cm}|p{0.7cm}|p{1cm}|p{0.7cm}|p{1cm}|p{0.7cm}|p{1cm}|}
\toprule
formula & {batch} & folder & {initial structures} & {volume per formula unit} & {formula units} & {jobs completed} & {converged without error} & {no ionic convergence} & {other VASP errors} & {mean number of steps to converge} & {maximum number of ionic steps} & {total computation time (hours)} & {cores} & {total core-hours} \\ \midrule
LiSi & 2 & Li1Si1\_02/ & 5000 & {320 \AA\textsubscript{3}} & 8 & 4999 & 4853 & 93 & 53 & 548.03 & 2000 & 5.18E+04 & 8 & 4.14E+05 \\ \hline
LiSi & 3 & Li1Si1\_03/ & 3000 & {400 \AA\textsubscript{3}} & 8 & 2998 & 2875 & 77 & 46 & 581.66 & 2000 & 6.38E+04 & 8 & 5.11E+05 \\ \hline
LiSi & 4 & Li1Si1\_04/ & 2000 & {560 \AA\textsubscript{3}} & 8 & 2000 & 1912 & 60 & 28 & 620.90 & 2000 & 2.89E+04 & 8 & 2.31E+05 \\ \hline
LiSi & 5 & Li1Si1\_05/ & 5000 & - & 1-16 & 4992 & 4680 & 260 & 52 & 558.11 & 2000 & 9.47E+04 & 8 & 7.57E+05 \\ \hline
Li2Si & 2 & Li2Si1\_02/ & 5000 & - & 1-12 & 4992 & 4596 & 343 & 53 & 614.15 & 2000 & 1.15E+05 & 8 & 9.21E+05 \\ \hline
Li2Si & 3 & Li2Si1\_03/ & 5000 & - & 2-4 & 5000 & 4898 & 5 & 97 & 317.56 & 2000 & 1.69E+04 & 8 & 1.35E+05 \\ \hline
Li7Si2 & 3 & Li7Si2\_03/ & 5000 & - & 1-4 & 4983 & 4515 & 412 & 56 & 685.26 & 2000 & 1.31E+05 & 8 & 1.05E+06 \\ \hline
Li7Si2 & 4 & Li7Si2\_04/ & 5000 & - & 4 & 4991 & 3980 & 975 & 36 & 960.60 & 2000 & 1.56E+05 & 16 & 2.50E+06 \\ \hline
Li7Si2 & 5 & Li7Si2\_05/ & 5000 & - & 4 & 4993 & 3972 & 967 & 54 & 950.25 & 2000 & 1.51E+05 & 16 & 2.42E+06 \\ \hline
Li13Si4 & 2 & Li13Si4\_02/ & 10000 & {200\%} & 2 & 9639 & 7572 & 1956 & 111 & 1,016.14 & 2000 & 7.06E+05 & 8 & 5.65E+06 \\ \hline
Li13Si4 & 3 & Li13Si4\_03/ & 5000 & - & 1-2 & 4981 & 4365 & 558 & 58 & 763.60 & 2000 & 1.61E+05 & 8 & 1.29E+06 \\ \hline
Li15Si4 & 2 & Li15Si4\_02/ & 5000 & - & 1-2 & 4916 & 4299 & 548 & 69 & 812.51 & 2000 & 1.90E+05 & 8 & 1.52E+06 \\ \hline
Li1O1 & 1 & Li1O1\_01/ & 5000 & - & 1-16 & 4221 & 2086 & 1798 & 337 & 399.45 & 2000 & 3.06E+05 & 8 & 2.45E+06 \\ \hline
Li2O & 1 & Li2O1\_01/ & 5000 & - & 1-12 & 4906 & 4590 & 141 & 175 & 474.18 & 2000 & 9.87E+04 & 8 & 7.90E+05 \\ \hline
SiO2 & 3 & SiO2\_03/ & 10000 & - & 1-12 & 9160 & 138 & 1091 & 7931 & 245.10 & 2000 & 1.79E+05 & 8 & 1.43E+06 \\ \hline
SiO2 & 4 & SiO2\_04/ & 10000 & - & 1-12 & 9620 & 152 & 1504 & 7964 & 308.62 & 2000 & 2.48E+05 & 8 & 1.99E+06 \\ \hline
SiO2 & 5 & SiO2\_05/ & 1000 & - & 2 & 992 & 443 & 535 & 14 & 1,033.90 & 2000 & 4.13E+04 & 32 & 1.32E+06 \\ \hline
SiO2 & 6 & SiO2\_06/ & 6000 & - & 2-4 & 5995 & 5238 & 597 & 160 & 1,019.71 & 3000 & 6.68E+04 & 32 & 2.14E+06 \\ \hline
SiO2 & 7 & SiO2\_07/ & 2000 & - & 6 & 1996 & 1384 & 600 & 12 & 1,492.28 & 3000 & 6.56E+04 & 32 & 2.10E+06 \\ \hline
SiO2 & 8 & SiO2\_08/ & 2000 & - & 8 & 1922 & 1102 & 792 & 28 & 1,619.18 & 3000 & 1.18E+05 & 32 & 3.78E+06 \\ \hline
SiO2 & 9 & SiO2\_09/ & 2000 & - & 12 & 1733 & 811 & 915 & 7 & 1,884.37 & 3000 & 1.92E+05 & 32 & 6.14E+06 \\ \hline
Li4SiO4 & 1 & Li4Si1O4\_01/ & 5000 & - & 2 & 4887 & 353 & 3276 & 1258 & 350.69 & 2000 & 2.64E+05 & 32 & 8.44E+06 \\ \hline
Li2Si2O5 & 1 & Li2Si2O5\_01/ & 5000 & - & 2 & 4979 & 3452 & 1431 & 96 & 1,549.92 & 3000 & 1.49E+05 & 32 & 4.75E+06 \\ \hline
Li15Si4 & {Li pseudopotential, conjugate algorithm} & Li15Si4\_02\_fa/ & 800 & - & 1-2 & 798 & 675 & 114 & 9 & 787.60 & 2000 & 1.52E+04 & 96 & 1.46E+06 \\ \hline
Li15Si4 & {k-point density(0.5)} & Li15Si4\_02\_k/ & 800 & - & 1-2 & 799 & 678 & 112 & 9 & 820.59 & 2000 & 4.07E+03 & 96 & 3.91E+05 \\ \hline
Li15Si4 & {k-point density(1)} & Li15Si4\_02\_k2/ & 800 & - & 1-2 & 800 & 690 & 100 & 10 & 853.69 & 2000 & 1.31E+03 & 96 & 1.25E+05 \\ \hline
Li15Si4 & {step size (0.8)} & Li15Si4\_02\_p/ & 800 & - & 1-2 & 800 & 781 & 8 & 11 & 525.45 & 2000 & 9.04E+03 & 96 & 8.68E+05 \\ \hline
TOTAL & {} &  & 116200 &  & {} & {} & {} & {} & {} & {} & {} & {} & {} & 5.56E+07 \\ \bottomrule

\end{tabular}%
}
\end{tiny}
\end{table}
\end{landscape}

\begin{table}[hbt!]
  \caption{{\bf Distribution of the random structure dataset}. The first two rows correspond to the components of force vectors and stress tensors. Max force magnitude refers to the maximum force magnitude in each structure. Max stress component is the maximum value of the stress tensor components ($\sigma_{xx}, \sigma_{yy}, \sigma_{zz}, \sigma_{xy}, \sigma_{yz}, \sigma_{zx}$) in each structure. }
  \label{data_details}
  \centering
  \begin{tabular}{lllll}
    \toprule             
     & mean    & standard deviation    & min   & max\\
    \midrule
    Force (eV/\AA)  &   0.000 & 0.208 & -4.797 & 4.797   \\
    Stress  (kBar) & 0.733 & 12.318 & -192.012 & 439.783   \\
    max force magnitude (eV/\AA) & 0.453 & 0.442 & 0.000 & 4.797   \\
    max stress component (kBar) & 8.662 & 23.555 & 0.000 & 439.783 \\
    \bottomrule
  \end{tabular}
\end{table}

\section{Hyperparameter search}
Hyperparameters were selected via grid search over the hyperparameters in table~\ref{hyper}. The number of epochs at which learning rates were decreased is tabulated for each value of Gaussian noise augmentation in table~\ref{lrd}.

\begin{table}[hbt!]
  \caption{{\bf Range of hyperparameters.}\\
  Cutoff radius: Cutoff radius for constructing edges between particles. We constructed edges between all pairs of particles located closer than this cutoff distance. \\
  *Size of MLP layers in MEGNet: n1, n2 and n3 are used as in ref. \cite{chen_graph_2019}.}
  \centering
  \begin{tabular}{ll} 
    \toprule             
    Hyperparameter    & Range \\
    \midrule
    Cutoff radius & 4\AA, 5\AA \\
    Number of message-passing layers & 3, 5, 7, 9\\
    Batch size & 32, 64, 128 \\
    Epochs for decreasing learning rates & [5,40], intervals of 5 \\
    n1, size of first MLP layer in MEGNet* & 32,64,128 \\
    n2, size of second MLP layer in MEGNet* & 32,64,128 \\
    n3, size of third MLP layer in MEGNet* & 32,64,128 \\
    \bottomrule
  \end{tabular}\label{hyper}
\end{table}

\begin{table}
  \caption{{\bf Learning rate decay.}\\
  Learning rates were decreased by a factor of 0.5 every N epochs tabulated below.}
  \centering
  \begin{tabular}{ll} 
    \toprule             
    Gaussian noise std (\AA) & N\\
    \midrule
    0   &15\\
    0.01    &20\\
    0.02    &20\\
    0.03    &25\\
    0.04    &30\\
    0.05    &30\\
    0.06    &30\\
    \bottomrule
  \end{tabular}\label{lrd}
\end{table}

\section{Energy predictions on random structure data}
The model trained on random structure energies achieved a MAE of 13.5 meV/atom. This is higher than other machine learning potentials on the Li-Si systems, such as \cite{artrith_constructing_2018, onat_implanted_2018, xu_deep-learning_2020}. As random structure data is quite different from what \cite{artrith_constructing_2018, onat_implanted_2018, xu_deep-learning_2020} are trained on, we trained and tested the same model on our MD data. The test MAE for MD energies is 3.8 meV/atom, similar to what's reported in these works.

\section{Relaxation algorithm affects noise in relaxation data}

\begin{figure}[hbt!]
  \centering
  \includegraphics[width=6cm]{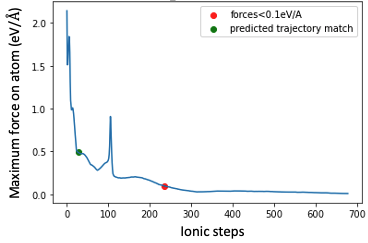}
  \caption{Sample trajectory of Li2Si in which GNN model did not reproduce the VASP relaxation trajectory. The red dot corresponds to where the maximum force on the atoms reach 0.1 eV/\AA and the simulation terminates. The green dot is the ionic step until the GNN simulation matches VASP. There are large spikes in atomic forces in this trajectory.}\label{trajectory}
\end{figure}

\begin{figure}[hbt!]
    
  \centering
  \includegraphics[width=8cm]{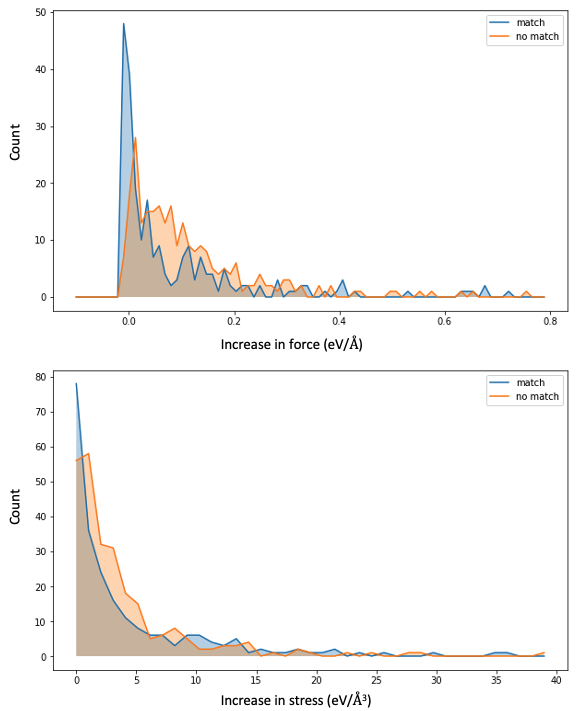}
  \caption{The distribution of upward spikes in relaxation trajectory for maximum forces on atoms (top) and maximum component of stress tensor (bottom). The trajectories where our graph network simulations match the VASP trajectory are plotted in blue(match), and the trajectories where our simulations do not match are plotted in yellow(no match). }\label{force_stress_spikes}
\end{figure}

We used the steepest descent algorithm as implemented in VASP (IBRION=3, SMASS=2, ISIF=3) and adjusted the time step parameter (POTIM) to a small value(0.4, default is 0.5) and checked all relaxations have a monotonic decrease in energy over ionic steps. But in some cases, we see large upward spikes in force or stress trajectory, as in figure \ref{trajectory}. To investigate whether these spikes introduce noise in our simulations, we quantified the amount of upward spikes in the relaxation trajectories that are successfully simulated by our model and those that fail. We calculated the maximum amount of increase in forces (maximum force applied to the atoms in the structure) and stresses (maximum component of the stress tensor) within 5 ionic steps for Li\textsubscript{2}Si structures in the test set. We plot the distribution of those spikes in figure \ref{force_stress_spikes}. We see that for both forces and stresses, the distribution of spikes is generally larger for the trajectories that our model fails to simulate correctly. The distribution suggests that some of our model's failures may attributed to unphysical behavior in DFT simulations with VASP used to construct the ground-truth data.

\section{Experimental structure of Li\textsubscript{15}Si\textsubscript{4}}
\begin{figure}[hbt!]
  \centering
  \includegraphics[width=6cm]{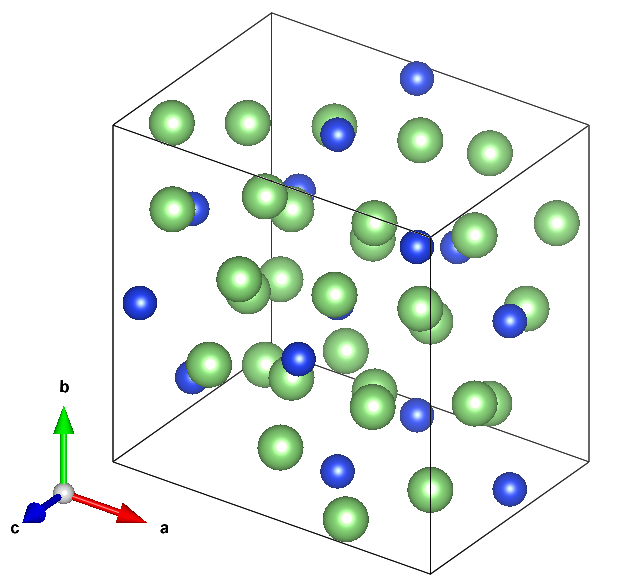}
  \includegraphics[width=6cm]{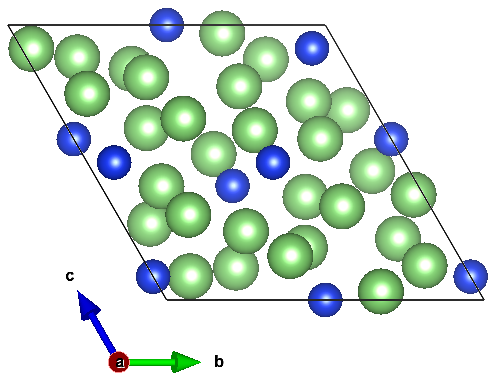}
  \caption{Experimental structure of Li\textsubscript{15}Si\textsubscript{4}}\label{li15si4}
\end{figure}

The experimental structure of Li\textsubscript{15}Si\textsubscript{4}\cite{dupke_structural_2012} is visualized in figure~\ref{li15si4}. 

\bibliography{main}
\bibliographystyle{icml2021}